\documentclass[final,1p,times]{elsarticle}

\usepackage[bookmarks, colorlinks=true, linktoc=page, pdftex, linkcolor=blue, citecolor=blue, urlcolor=blue]{hyperref}

\usepackage{subfigure}
\usepackage{graphicx}

\usepackage{textcomp}
\usepackage{amsmath,amssymb,amsthm}
\usepackage[euler]{textgreek}

\usepackage{lineno}
\usepackage[colorinlistoftodos,prependcaption,textsize=tiny]{todonotes}

\begin{document}


\begin{frontmatter}
\journal{Nucl. Instr. Meth. A}

\title{ARRAY: An Open Source, Modular and Probe-Card based System with Integrated Switching Matrix for Characterisation of Large Area Silicon Pad Sensors}


\author[cern-ep]{Erica~Brondolin}
\author[cern-ep]{Dominik~Dannheim}
\author[cern-ep]{Szymon~Kulis}
\author[cern-ep]{Andreas~A.~Maier}
\author[cern-ep,tu]{Florian~Pitters\fnref{hephy}\corref{cor}}
\author[cern-ep]{Thorben~Quast}
\author[cern-ep]{Eva~Sicking}
\address[cern-ep]{CERN, Geneva, Switzerland}
\address[tu]{TU Wien, Vienna, Austria}

\fntext[hephy]{Now at HEPHY, Vienna, Austria}
\cortext[cor]{Corresponding author: florian.pitters@pm.me}

\begin{abstract}
Silicon pad sensors are proposed as active material in highly granular sampling calorimeters of future collider experiments such as the Compact Linear Collider (CLIC) or the International Linear Collider (ILC).
The electromagnetic section of these designs often include O(1000\,m$^2$) of silicon pad sensors.
For the luminosity measurement, a dedicated forward calorimeter called LumiCal is foreseen.
More recently, the CMS experiment has decided to use the same concept in its endcap calorimeter upgrade for the HL-LHC.
The sensors are typically produced from 6- or 8-inch wafers and consist of a few hundred smaller cells, each with an area of O(0.1 to 1\,$\text{cm}^2$).
For the prototyping phase of these projects, several design choices have to be evaluated while for mass production, thousands of sensors have to be tested for quality control.
For the electrical characterisation of these sensors, it is important to bias them under realistic conditions.
To fulfil these requirements, ARRAY, a compact, modular and cost efficient system for large area silicon pad sensor characterisation has been developed and successfully commissioned.
It consists of two plugin printed circuit boards: an active switching matrix with 512 input channels that holds all controls and a passive probe card that connects to the sensor.
The latter can then be adapted to any sensor geometry.
All design files are open source.
The system has been used to measure currents ranging from 500\,pA to 5\,\textmu A and capacitances between 5\,pF and 100\,pF.
A precision of better than 0.2\,pF on capacitance measurements in that range can be achieved.
Examples of calibration and measurement results for leakage current and capacitance are presented.
\end{abstract}


\end{frontmatter}


\tableofcontents


\section{Introduction}
\label{sec:introduction}

Imaging calorimetry systems optimised for Particle Flow Analysis \cite{Thomson:2009rp} are proposed for future collider experiments such as the Compact Linear Collider (CLIC)~\cite{ref:clic_summary} or the International Linear Collider (ILC)~\cite{Behnke:2013xla}.
The electromagnetic section of these designs often includes O(1000\,m$^2$) of silicon pad sensors.
For the luminosity measurement at such colliders, a dedicated forward calorimeter called LumiCal~\cite{Abramowicz:2010bg} with similar structure has been proposed to measure Bhabha scattering at small angles.
LumiCal is developed within the Forward Calorimetry collaboration (FCAL).
More recently, the Compact Muon Solenoid experiment (CMS) has decided to use the same concept of a highly granular sampling calorimeter in the upgrade of its endcap calorimeters~\cite{hgcal_tdr} for the High Luminosity phase at the Large Hadron Collider (HL-LHC).
This upgrade, commonly called HGCAL (High Granularity Calorimeter), will use about 27000 silicon pad sensors to cover roughly 600\,m$^2$.
Typically, such sensors are made from 6- or 8-inch wafers, divided into several hundred DC-coupled pads of O(0.1 to 1\,$\text{cm}^2$) in size.

For the prototyping phase, several design choices concerning the sensors and their implications on other parts of the system, e.g. the front-end electronics, are investigated.
For mass production, thousands of sensors will have to be characterised for quality control.
It is therefore essential to establish a fast and automatic testing procedure.
Important characteristics to be measured include the current versus voltage (IV) and capacitance versus voltage (CV) behaviour.
From such measurements, parameters like the breakdown voltage V$_{\text{bd}}$ or the capacitance at full depletion C$_{\text{fd}}$ can be extracted.
As these parameters depend on the exact electric field configuration inside the sensor, it is essential that all pads are biased during sensor characterisation, similar to the operating of the experiment.
This is especially true for the guard ring area that surrounds the pad matrix and is generally the most prone for high-field regions causing breakdown.
Due to the lack of a common bias structure in the sensors, a probe-card based system that connects to all pads on the sensor simultaneously is required.

The testing setup should be capable of measuring multiple sensor layouts, sustain bias voltages of $\pm$1000\,V and handle up to 500 input channels, approximately corresponding to the number of 0.5\,$\text{cm}^2$ pads on an 8-inch wafer.
Currents ranging from hundreds of pA to 10\,\textmu A and capacitances up to 100\,pF have to be measured accurately and in a reasonable measurement time.
Ideally, the system would also be compact, cheap and easily portable to different laboratories.
To fulfil these requirements, a solution with a multi-channel multiplexer and one set of precision instruments was selected.
Commercially available instruments were selected for CV and IV measurements.
For the switching system however, no suitable commercial solution was identified.
While several commercial switching systems offer the required precision and long term stability, the number of channels is limited.
Therefore, one would have to stack several instruments in order to support the required number of channels.
An additional disadvantage of this type of systems is related to the input connectivity requiring hundreds of high quality shielded cables to be connected to the probe card.
Due to limitations of commercial systems, a dedicated switching matrix together with a probe card has been developed  to fulfil all requirements of performance, compactness and cost.
The system is named ARRAY (switching mAtRix pRobe cArd sYstem).

Section~\ref{sec:setup} gives a technical description of the system. Section~\ref{sec:results} shows results from the system's calibration as well as example results from IV and CV measurements on prototype sensors for HGCAL and LumiCal. Section~\ref{sec:discussion} discusses the measurement range of the system, the precision of the capacitance measurement and the special case of measuring irradiated sensors. Finally, section~\ref{sec:summary} gives a summary.


\section{The ARRAY Sensor Testing System}
\label{sec:setup}

\begin{figure}[htb]
\begin{center}
    \subfigure[]{
        \includegraphics[height=0.30\textheight]{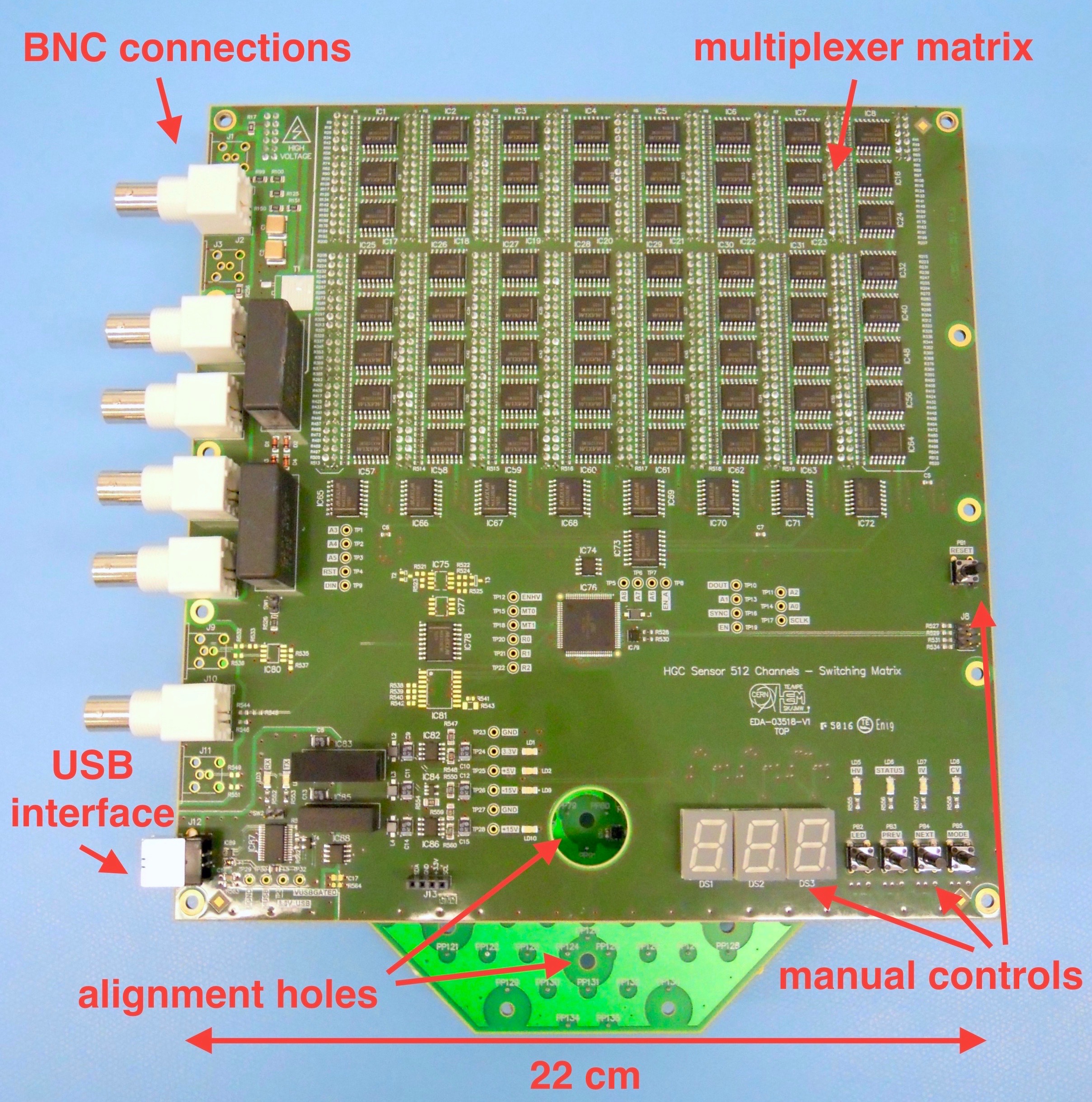}
        \label{fig:assembly:a}}%
    \subfigure[]{
        \includegraphics[height=0.30\textheight]{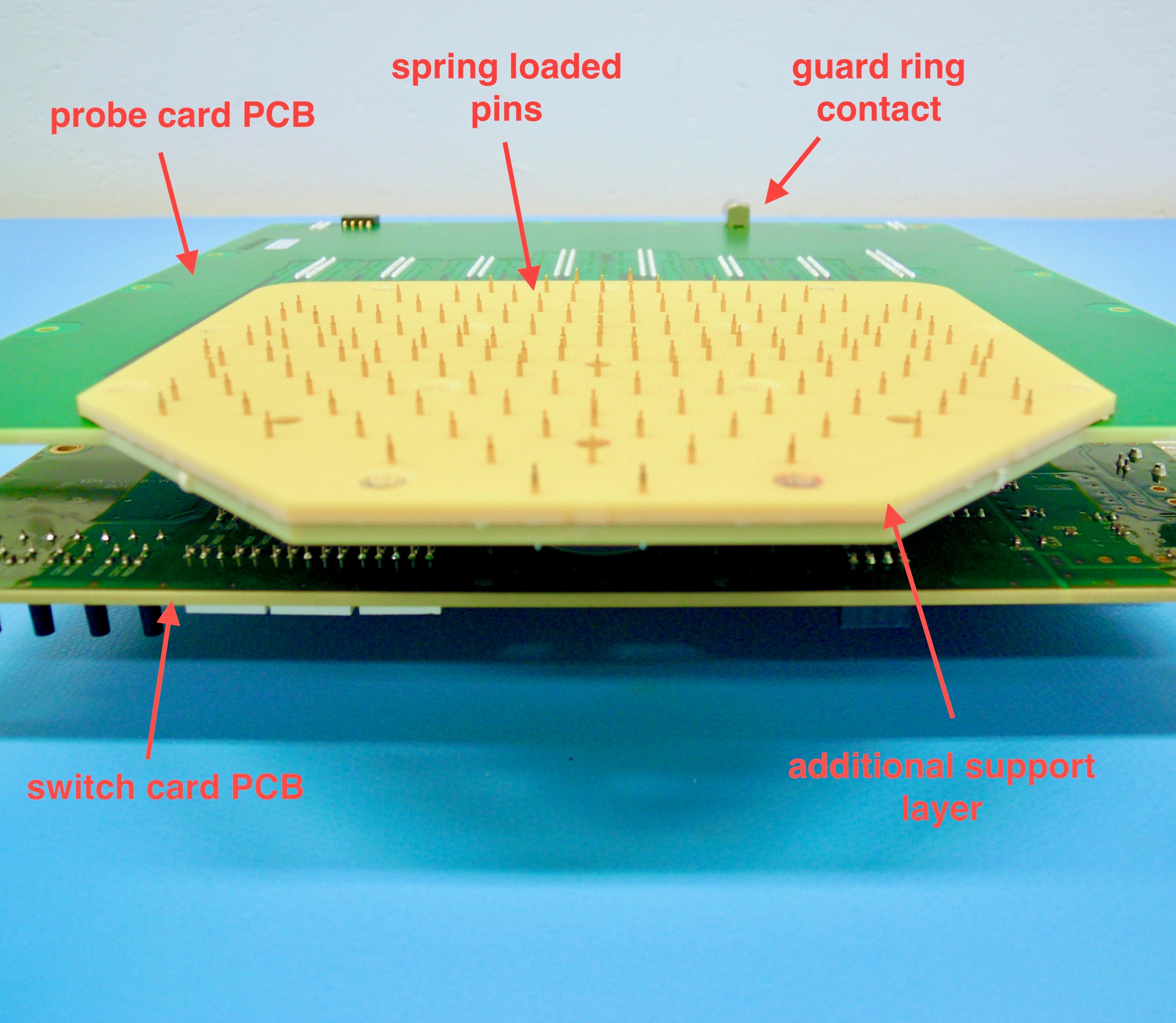}
        \label{fig:assembly:b}}
    \caption{Picture of the assembled two card system with view from (a) the top and (b) the bottom.}
    \label{fig:assembly}
\end{center}
\end{figure}

The switching matrix is designed as a plug-in printed circuit board (PCB), called switch card, that sits directly on top of the probe card.
Fig.~\ref{fig:assembly} shows the system from (a)~the top and (b)~the bottom.
The switch card is essentially a large array of multiplexers that controls the measurement while the probe card is a passive device that provides the connectivity to the sensor.
As the sensor geometry changes, only the probe card layout has to be adapted.
So far, six cards have been designed for the CMS and FCAL collaborations.
Spring-loaded pins are used to connect the probe card to sensor cells with $\sim$\,1\,mm$^2$ contact pads.
The probe card can also be equipped with finer probe tips if a more precise connection should be needed for e.g. silicon strip sensors.



The two cards are mechanically attached to a probe station, which itself is inside a light-tight box.
The sensor is held in place by a vacuum chuck.
Viewing holes inside the cards allow for x/y alignment of the probe card relative to the sensor before contacting.

\begin{figure}[htb]
\begin{center}
    \includegraphics[width=0.90\textwidth]{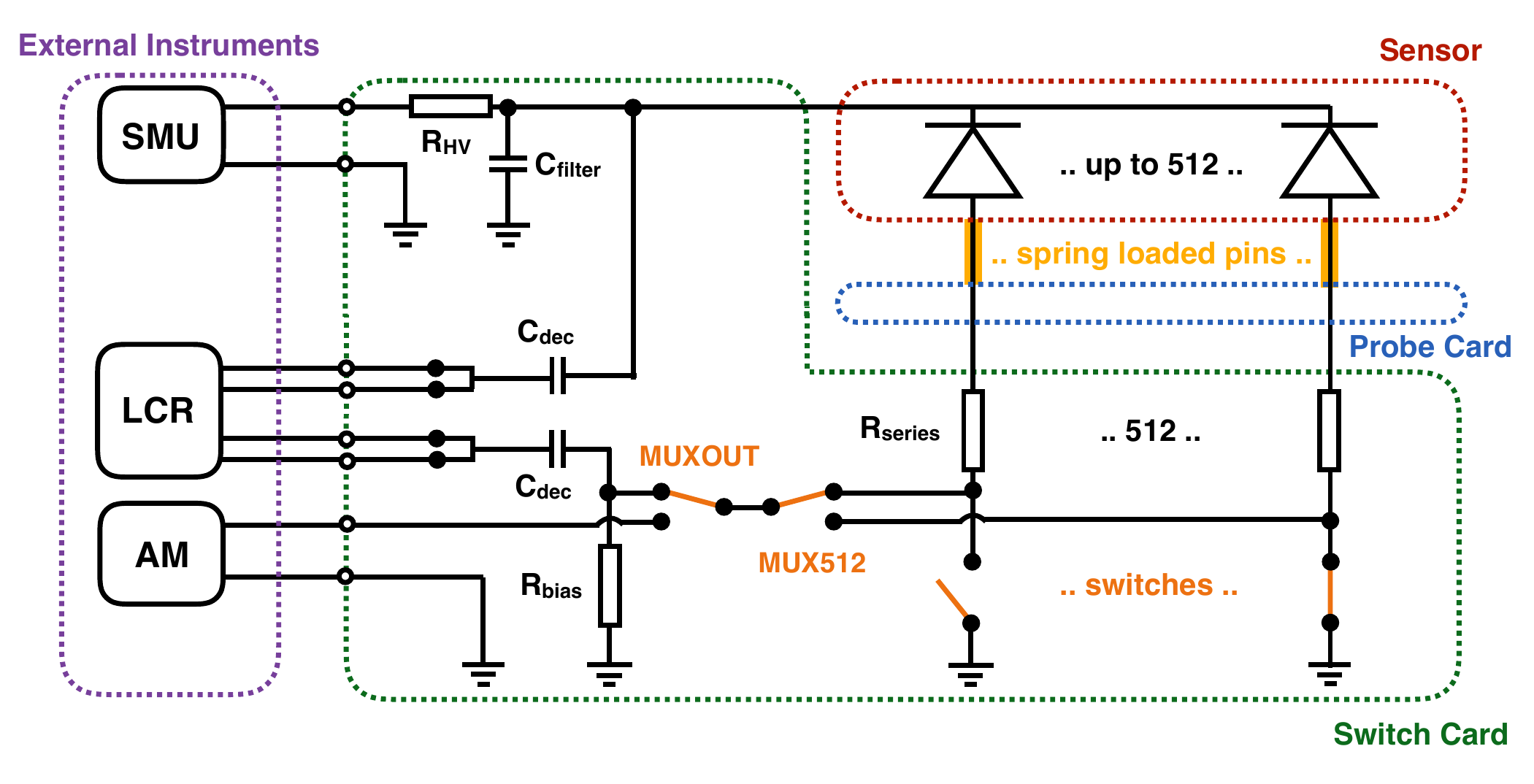}
    \caption{A simplified circuit diagram of the system. For CV measurements, the pad-under-test returns to ground via a variable bias resistor, and for IV measurements via the ammeter. All other pads are directly shorted to ground. A hard-coded switching scheme guarantees that all pads are biased at all times. Acronyms are introduced in Sec.~~\ref{sec:switch_card}.}
    \label{fig:schematic}
\end{center}
\end{figure}

A simplified circuit diagram of the full system is shown in Fig.~\ref{fig:schematic}.
Infrastructure for IV and CV measurements, such as a high voltage filter circuit or decoupling capacitors, is implemented on the switch card.
From the user's side, only the HV power supply as well as instruments for current and capacitance measurements have to be provided.
Firmware and design files are available under the CERN Open Hardware license and can be found in~\cite{ref:repo_owhr}.
The production cost of the switch card are around 2500\,CHF per PCB.\footnote{Prices refer to a production batch of five units including components, assembly, tooling and production in Switzerland in 2018.}
The cost of the probe card is typically about a factor two less than for the switch card but depends on the required number of spring loaded pins.
The full system is roughly the size of an A4 paper stack.

\subsection{Probe Card}
\label{sec:probe_card}
The probe card consists of an array of spring-loaded pins with 1.4\,mm travel range that are used to contact all individual sensor pads (see Fig.~\ref{fig:assembly:b}).
The travel range avoids damaging the pads during contact and ensures a uniform contact over the full sensor area.
The chosen pins are 4.8\,mm long and have a rounded tip with a radius of 250\,\textmu m.
The contact marks on the sensor have about twice the diameter of the marks from a conventional 12\,\textmu m tip prober needle.
To enable a good electrical contact to the aluminium pads on the sensor, a force of about 25\,g per pin has to be applied.
This force has to be applied via the mechanical integration into the probe station.
Due to the spring-loaded nature of the pins, the positioning accuracy is limited to about $\pm$\,200\,\textmu m.
For the guard ring with typical contact pads of few tens of \textmu m, this precision is not sufficient.
Therefore, the probe card provides an additional input that allows the use of an external probehead for the guard ring connection with needles.

The pins, as well as the guard ring input, are then routed to the connectors on the back side of the probe card, which plug into the switch card.
The routing avoids parallel tracks and maximises clearance in order to limit parasitic capacitances.
Test capacitances are added to the probe card to allow for system calibration and cross checks.
Humidity and temperature sensors are integrated into the card to monitor environmental conditions.
For different sensor geometries, different probe cards are used, yet all cards are based on the same concept.
The probe card is mechanically fixed to a steel mounting frame.
The base PCB is 2.2\,mm thick.
In the contact area of the pins, an additional support layer of 2\,mm thickness is used to provide mechanical support and to facilitate the pin placement.
An array of standoffs between probe card and mounting frame prevents any significant bow in the PCB.
Typical flatness values are well below the travel of the pins.

\subsection{Switch Card}
\label{sec:switch_card}
The signal originating from the probe card is routed via a multiplexer (MUX512) towards the readout instruments.
To make the system compact and ensure a long lifetime, solid state multiplexers are used.
Due to the limited number of channels available in such devices and in order to minimise the parasitic capacitance, the MUX512 is implemented in a hierarchical way, with three levels of 8-to-1 multiplexers.
The output multiplexer (MUXOUT) enables switching between the IV and CV output circuits.
Multiplexers of type MAX328~\cite{ref:max328} have been selected due to their small series resistance of typically  1.5\,k\textOmega~and their leakage current of only a few pA.
An additional 1\,k\textOmega\:resistor offers protection to the multiplexers on the switch card should the potential on the low side of the diode drift too far away from ground.
The total value of R$_\text{series}$ from the multiplexers and this resistor is about 7\,k\textOmega.
Additionally, each channel can be shorted to ground using solid state switches of type ADG1414~\cite{ref:adg1414} to ensure proper bias for cells which are not being measured.

The switch card provides six coaxial BNC connections to a source measurement unit (SMU), LCR meter (LCR) and ammeter (AM).
Alternative triaxial connectors are also available for SMU and AM.
For the IV circuit, the output of MUXOUT is connected directly to the measurement device and then returned to ground.
There is also a simple ammeter implementation directly on the switch card that can be used for quick pass/fail qualification in mass testing.
In the case of the CV circuit, a high value resistor R$_\text{bias}$ to ground is added to ensure proper bias.
This resistor raises the impedance in the parallel circuit and allows for a measurement of the capacitance of a single pad, rather than all pads on the sensor.
Its value can be set to eight discrete values between 100\,k\textOmega~and 100\,M\textOmega.
Two 1\,\textmu F capacitors C$_\text{dec}$ decouple the LCR meter for the DC voltage.

The high voltage (HV) is connected to the switch card, routed via the filter network of R$_\text{HV}$ and C$_\text{filter}$ to the probe card and applied to the sensor back side via the chuck.
On the switch card, a microprocessor controls on-board components and provides the user interface.
The settings of the switch card can be changed manually (using buttons) or remotely via a software interface.
The switching scheme is hard-coded to always short all pads to ground first before switching the pad-under-test.
This way, switching under large loads is avoided.
Since the resistance of the diodes is typically well above 1\,G\textOmega, the change in bias voltage is negligible.

\subsection{Data Acquisition}
The switch card is controlled via a USB serial interface.
Two libraries have been developed to interface the card, one in Python and one in LabVIEW\textsuperscript{TM}.
They can be found in the design repository~\cite{ref:repo_owhr}.
These libraries also provide the data acquisition and monitoring with controls for a few common measurement instruments.

The total switching time between channels is around 100\,ms.
The measurement of the current or capacitance per channel takes between 1\,s and 2\,s, depending on the external devices and integration times that are used.
An additional delay of a few seconds is used per voltage step to account for the time constant introduced by C$_\text{dec}$ and R$_\text{bias}$.
In the authors' setup, a typical IV or CV scan for a sensor with about 150 pads takes about one hour for 15 voltage steps.
This is at least an order of magnitude faster than a semi-automatic probe station with multiple probe heads could achieve, due to the needed settling time after a change in voltage.



\section{Measurement Results}
\label{sec:results}

\subsection{System Calibration}

\begin{figure}[htb]
    \centering
    \includegraphics[width=0.80\textwidth]{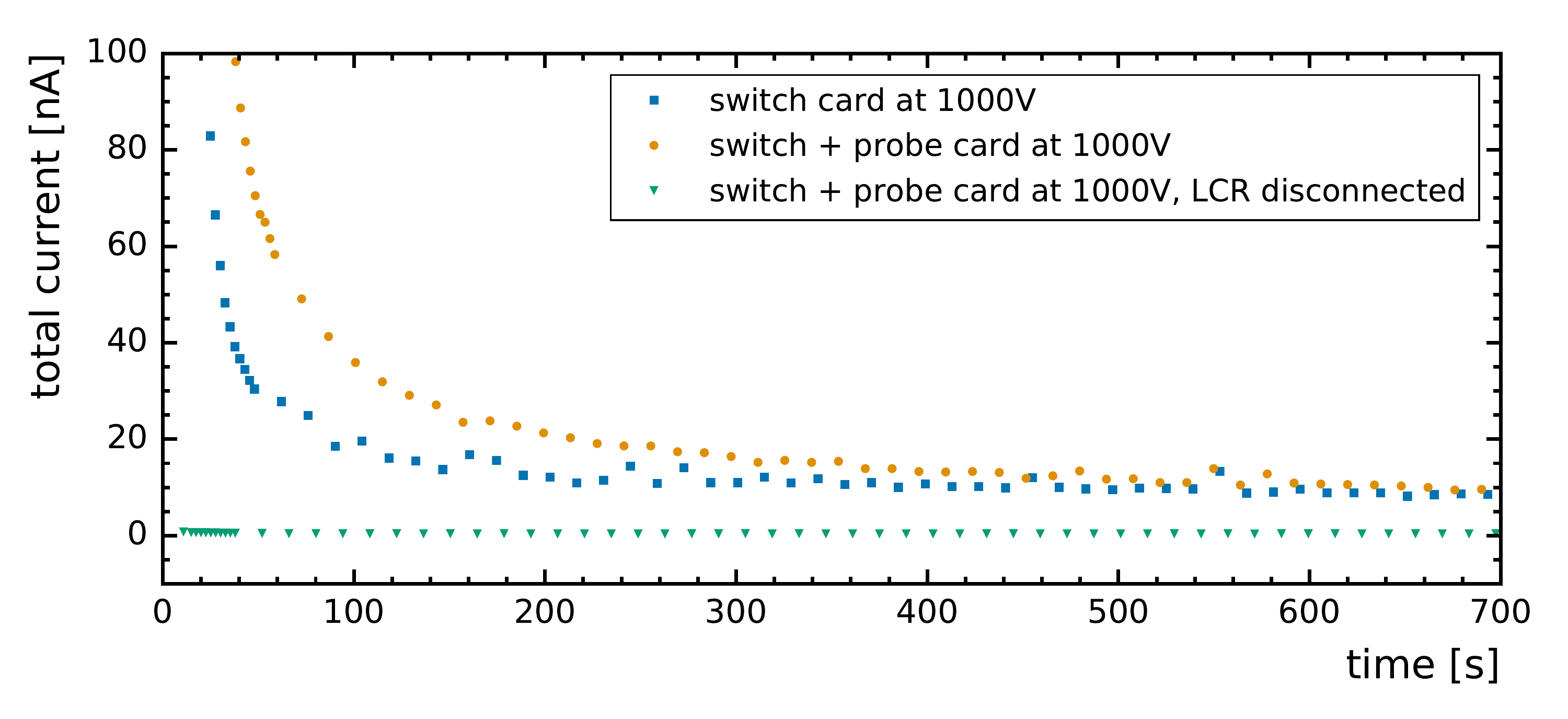}
    \caption{The system's total current over time at 1000\,V.}
    \label{fig:tot_bare_card}
\end{figure}

The total current of the system drawn from a high voltage power supply (here Keithley 2410) at 1000\,V as a function of time is shown in Fig.~\ref{fig:tot_bare_card} for the bare switch card, the assembly of switch card and probe card as well as the assembled system when the LCR meter is disconnected.
After charging of the capacitors, a small current of about 10\,nA remains for the bare and assembled two-card system.
This current disappears if the LCR meter is disconnected and can therefore be attributed to the leakage current of the C$_\text{dec}$ on the switch card. These measurements as well as all other measurements in this work have been recorded at room temperature and a humidity of about 50\%.

\begin{figure}[htb]
    \centering
    \includegraphics[width=0.80\textwidth]{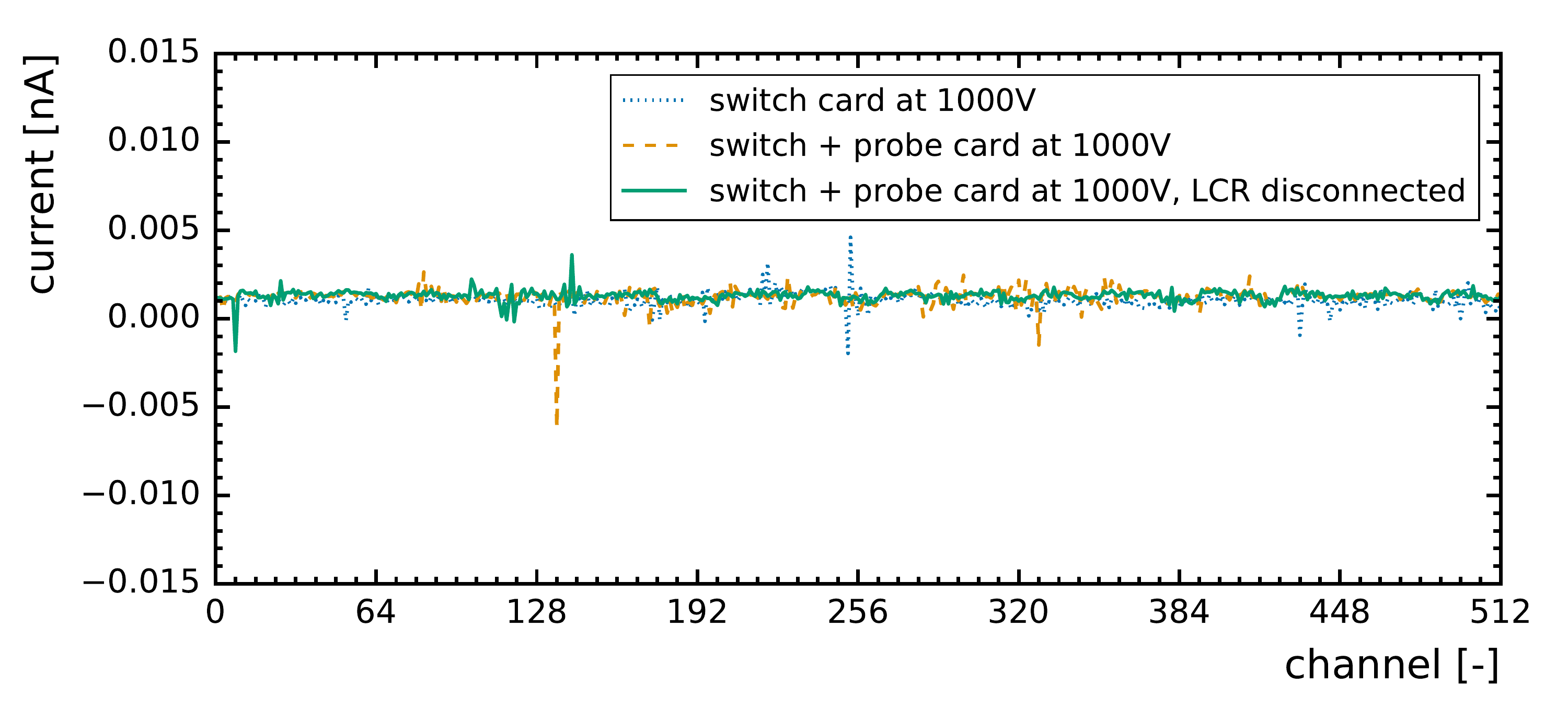}
    \caption{The system's leakage current per channel at 1000\,V.}
    \label{fig:iv_bare_card}
\end{figure}

The per-cell current is measured with an external ammeter (here Keithley 6487).
The leakage current per channel is less than 10\,pA at 1000\,V and is shown in Fig.~\ref{fig:iv_bare_card}.

\begin{figure}[htb]
    \centering
    \includegraphics[width=0.80\textwidth]{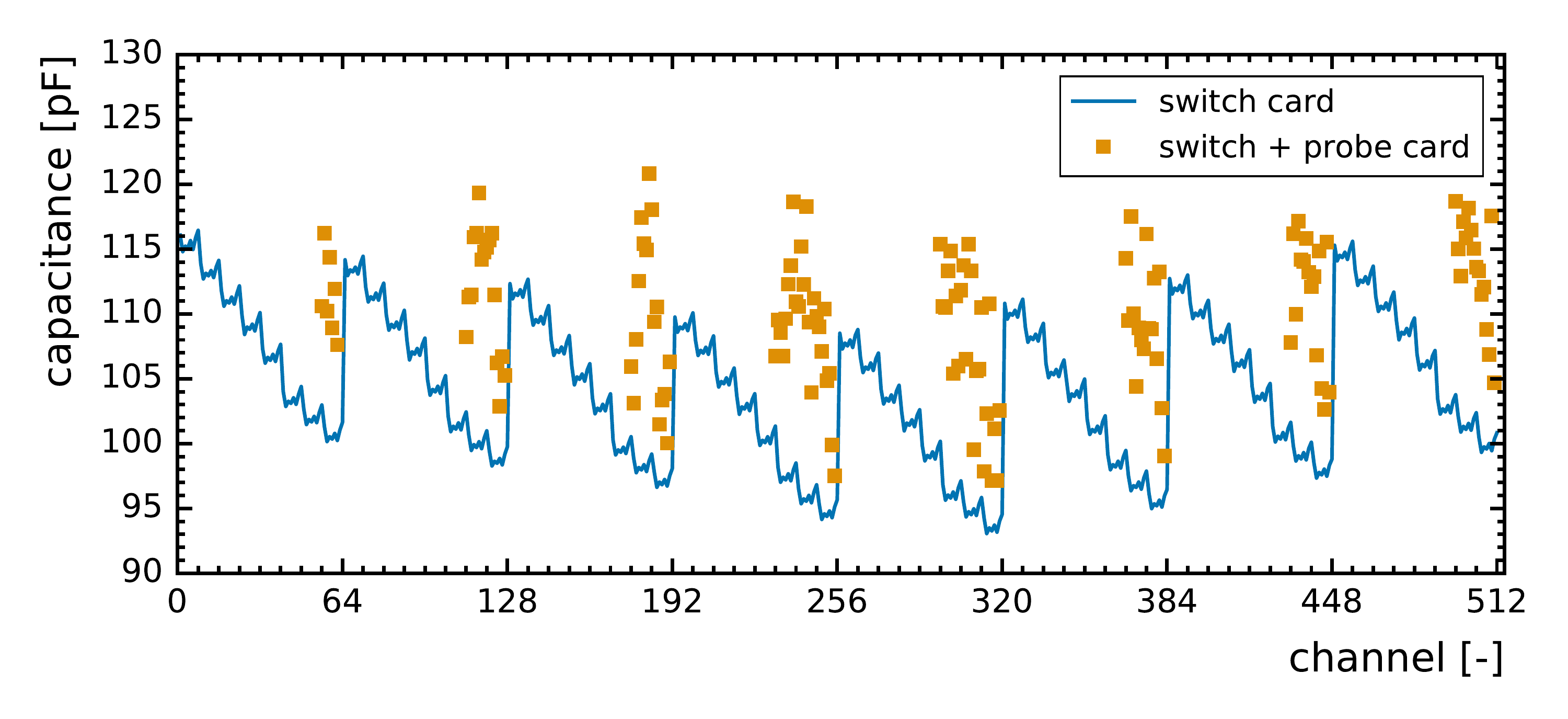}
    \caption{The capacitance per channel for the switch card with and without a 136 channel probe card connected. In the latter case, only the connected channels are measured. The measured values are then subtracted from consequent sensor measurements.}
    \label{fig:cv_alone}
\end{figure}

The capacitance is measured with an external LCR meter (here Keysight E4980A) and a parallel equivalent circuit is assumed.
Fig.~\ref{fig:cv_alone} shows the capacitance of the bare switch card and after assembly of switch and probe card.
For the bare switch card (solid line in Fig.~\ref{fig:cv_alone}), a step structure and two sub structures, each with a period of 8, are visible.
These structures are due to the three layers of multiplexers.
The absolute values can change by O(10\%) from PCB to PCB, however the structure remains unchanged.
Once the probe card is attached to the switch card, the capacitance of the channels with a connected trace on the probe card is increased by up to 20\,pF, depending on the routing of that trace.
These channels are shown as individual data points in Fig.~\ref{fig:cv_alone} for a probe card with 136 channels.
The capacitance values are constant over time and temperature and can therefore be corrected.

For sensor measurements, the corrected capacitance is obtained by subtracting the values of the open measurement shown in Fig.~\ref{fig:cv_alone},
\begin{equation}
     C_\text{cor} = C_\text{meas} - C_\text{open}.
     \label{eq:comp}
\end{equation}
This is known as open compensation.
Here, C$_\text{meas}$ is the capacitance of the ARRAY system plus sensor and C$_\text{open}$ the capacitance of the  system without sensor.
The latter is measured with all pins lifted from the sensor.
This is however not the exact compensation as this would require the removal of the single capacitance under test.
With a full-size silicon pad sensor, this is not possible and all pads are disconnected.
The systematic uncertainty introduced by this procedure has been analysed with a SPICE simulation and is less than 0.1\,pF in the frequency range between 1 and 100\,kHz.
An open/short compensation~\cite{keysight} has also been investigated in simulation and measurement but did not yield superior results.
All shown capacitance measurements in this work are corrected by open compensation.

\subsection{Example Results}
\begin{figure}[htb]
    \centering
    \subfigure[]{
        \includegraphics[height=0.26\textheight]{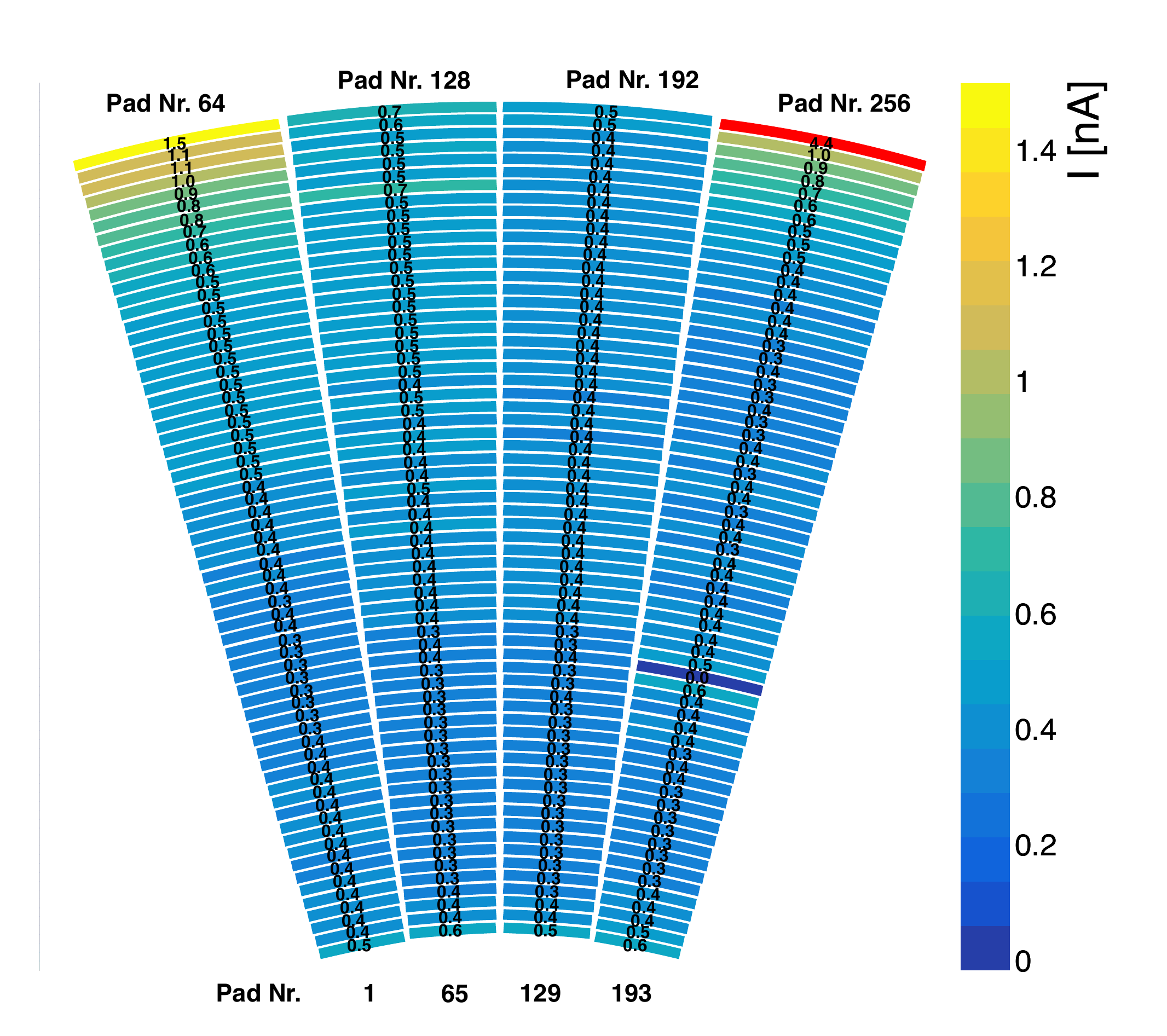}
        \label{fig:iv_map_lumi}}%
    \subfigure[]{
        \includegraphics[height=0.26\textheight]{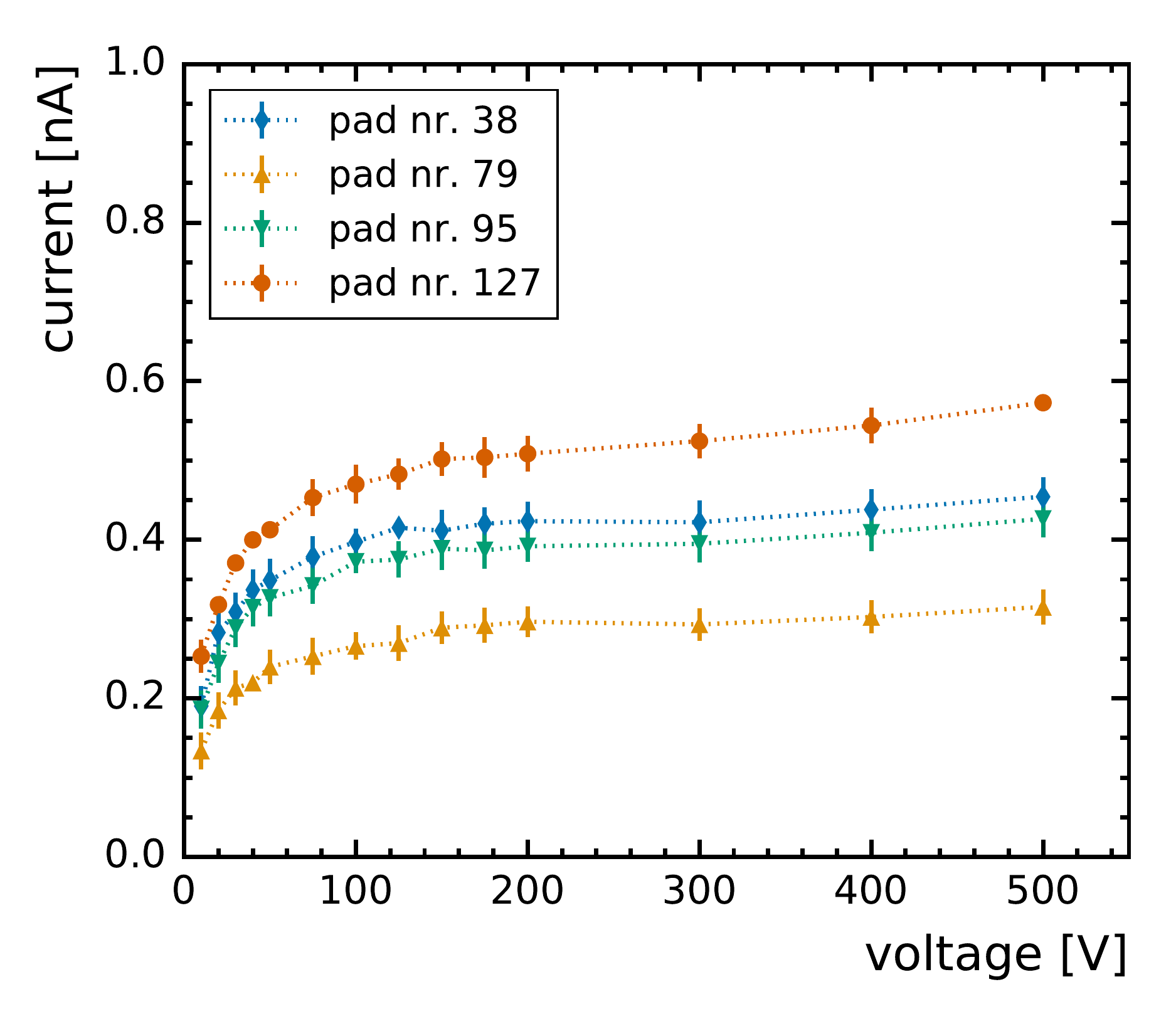}
        \label{fig:iv_curve_lumi}}
    \caption{(a) Leakage current map of a LumiCal protoype sensor at 500\,V bias (colour on-line). The current of one badly contacted pad is distributed to neighbouring pads. (b) Example IV curves from the same sensor. The error bars corresponds to the standard deviation from five consecutive measurements.}
    \label{fig:iv_lumi}
\end{figure}

This section shows example results from measurements on prototype sensors with the system.
Fig.~\ref{fig:iv_map_lumi} shows the leakage current at 500\,V from a 6-inch LumiCal prototype sensor.
The sensor consists of 256 p-on-n pads of varying sizes between about 16\,mm$^2$ to 40\,mm$^2$.
The thickness is 320\,\textmu m.
Here, one cell in the rightmost column has not been contacted properly and its current is distributed among neighbouring pads.
Obtaining a proper electrical contact for a few hundred pads with the spring-loaded pins requires a careful mechanical integration of the system into a probe station, to ensure a high level of parallelism between the chuck and the mechanical frame that hosts the two cards.
A forward bias test can quickly validate that good contact is established to all pads.
With some practise, this can typically be achieved on the first or second attempt.
Fig.~\ref{fig:iv_curve_lumi} shows a few example IV curve from the same sensor.
The error bars corresponds to the standard deviation from five consecutive measurements.


\begin{figure*}[htb]
    \centering
    \subfigure[]{
        \includegraphics[width=0.46\textwidth]{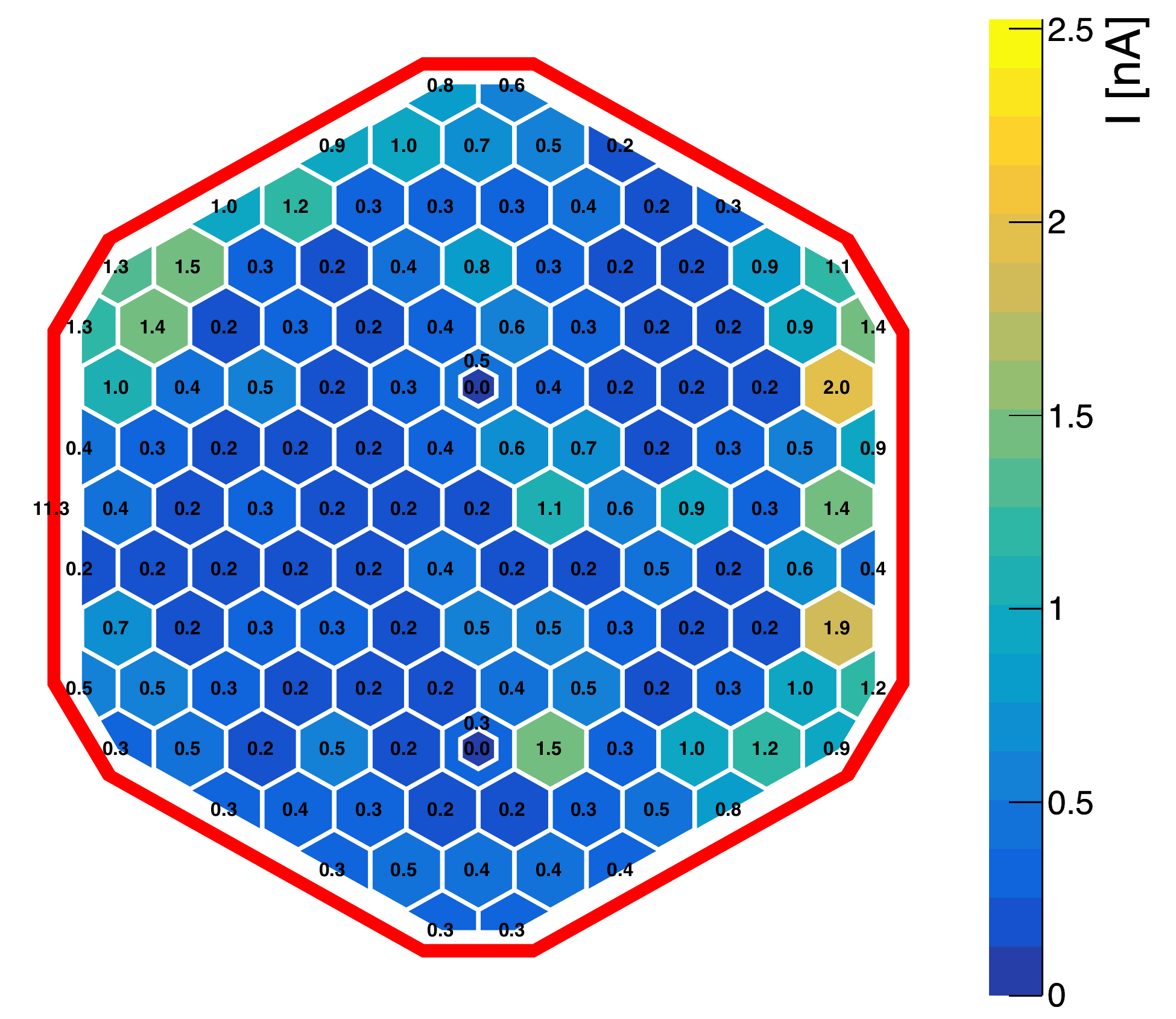}
        \label{fig:iv_map}}%
    \subfigure[]{
        \includegraphics[width=0.46\textwidth]{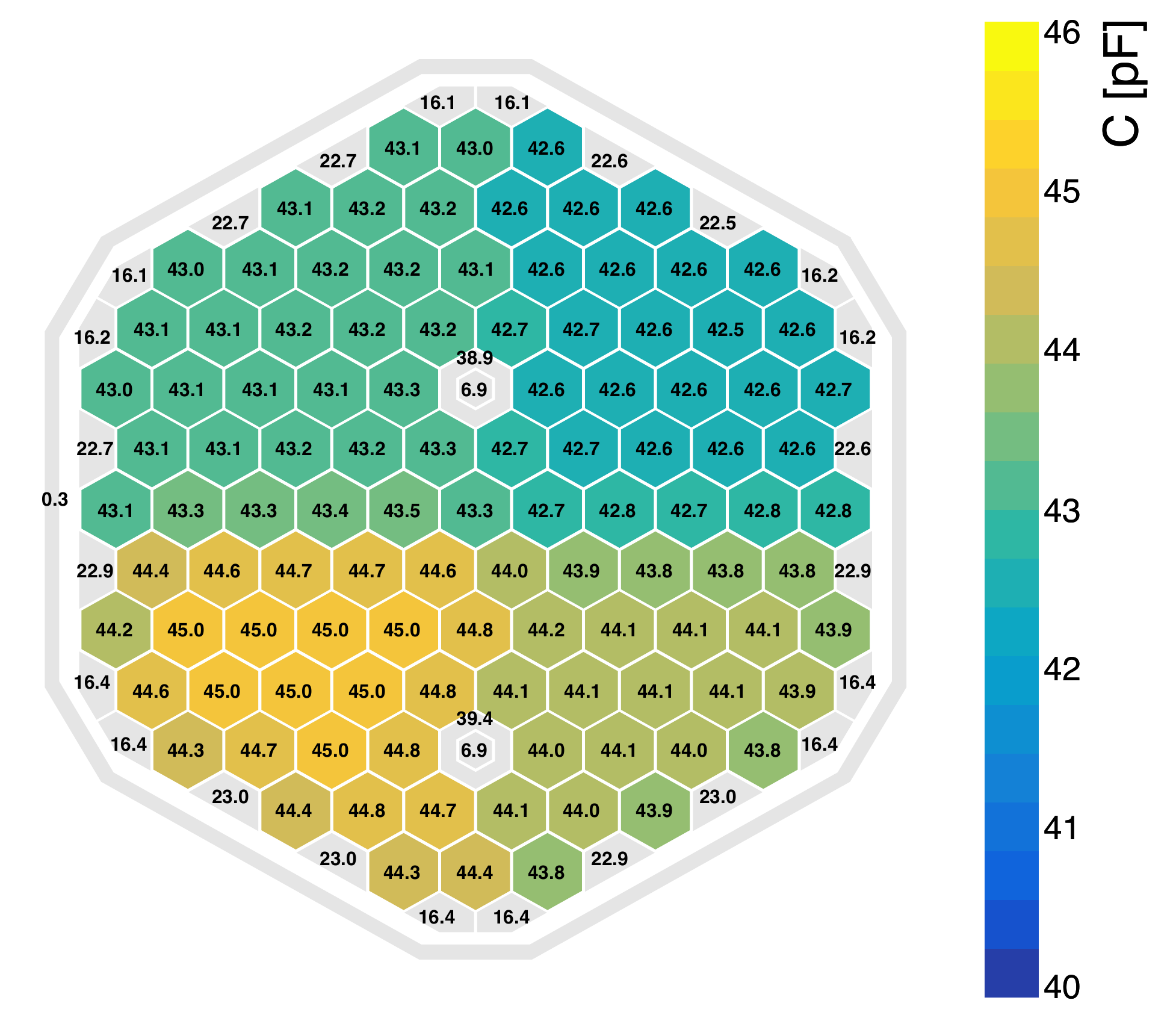}
        \label{fig:cv_map}}
    \caption{Pad values for (a) leakage current at 1000\,V and (b) capacitance at 400\,V measured with 1\,kHz for an example 6-inch 135 pad p-on-n sensor measured with the ARRAY system (colour on-line). The four quadrants in the capacitance measurements correspond to four different cell geometries on the sensor. The guard ring is depicted as a line surrounding the sensor pads.}
    \label{fig:maps}
\end{figure*}

Fig.~\ref{fig:iv_map} shows the leakage current of a prototype 6-inch p-on-n sensor for CMS HGCAL at 1000\,V.
The sensor consists of 135 mostly hexagonal pads of about 1\,$\text{cm}^2$ size surrounded by a grounded guard ring that is shown as a line around the sensor pads.
The sensor's nominal active thickness is 300\,\textmu m.
When comparing the total current measured at the SMU with the sum of all measured channels at 1000\,V, the results agree within a few nA (not shown).
This difference is caused by the small leakage current through the HV decoupling capacitors, as shown also in Fig.~\ref{fig:tot_bare_card}.
For R\&D purposes, the sensor is segmented into four quadrant geometries with different inter-pad gap distances, ranging from 20\,\textmu m to 80\,\textmu m.
Fig.~\ref{fig:cv_map} shows the capacitance at 400\,V (100\,V above full depletion) measured at 1\,kHz and R$_\text{bias}$ set to 10\,M\textOmega.
The influence of the different quadrants on the inter-pad capacitance and therefore the pad capacitance can be resolved.
The results match the expectations from prior TCAD simulations~\cite[pg. 119]{hgcal_tdr} at the nominal thickness within $\pm$0.5\,pF.

\begin{figure}[htb]
    \centering
    \subfigure[]{
        \includegraphics[width=0.49\textwidth]{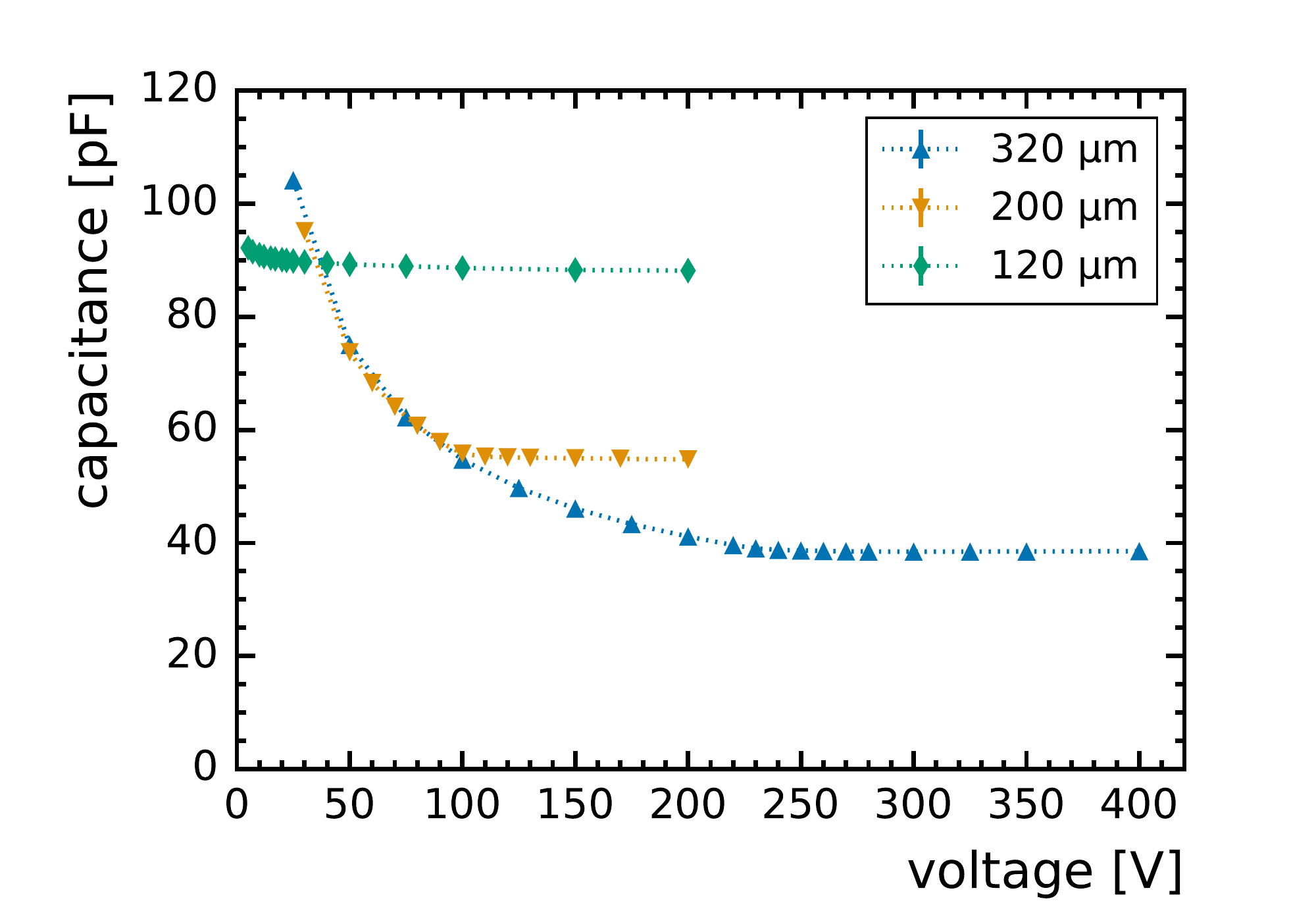}
        \label{fig:cv_curve}}%
    \subfigure[]{
        \includegraphics[width=0.49\textwidth]{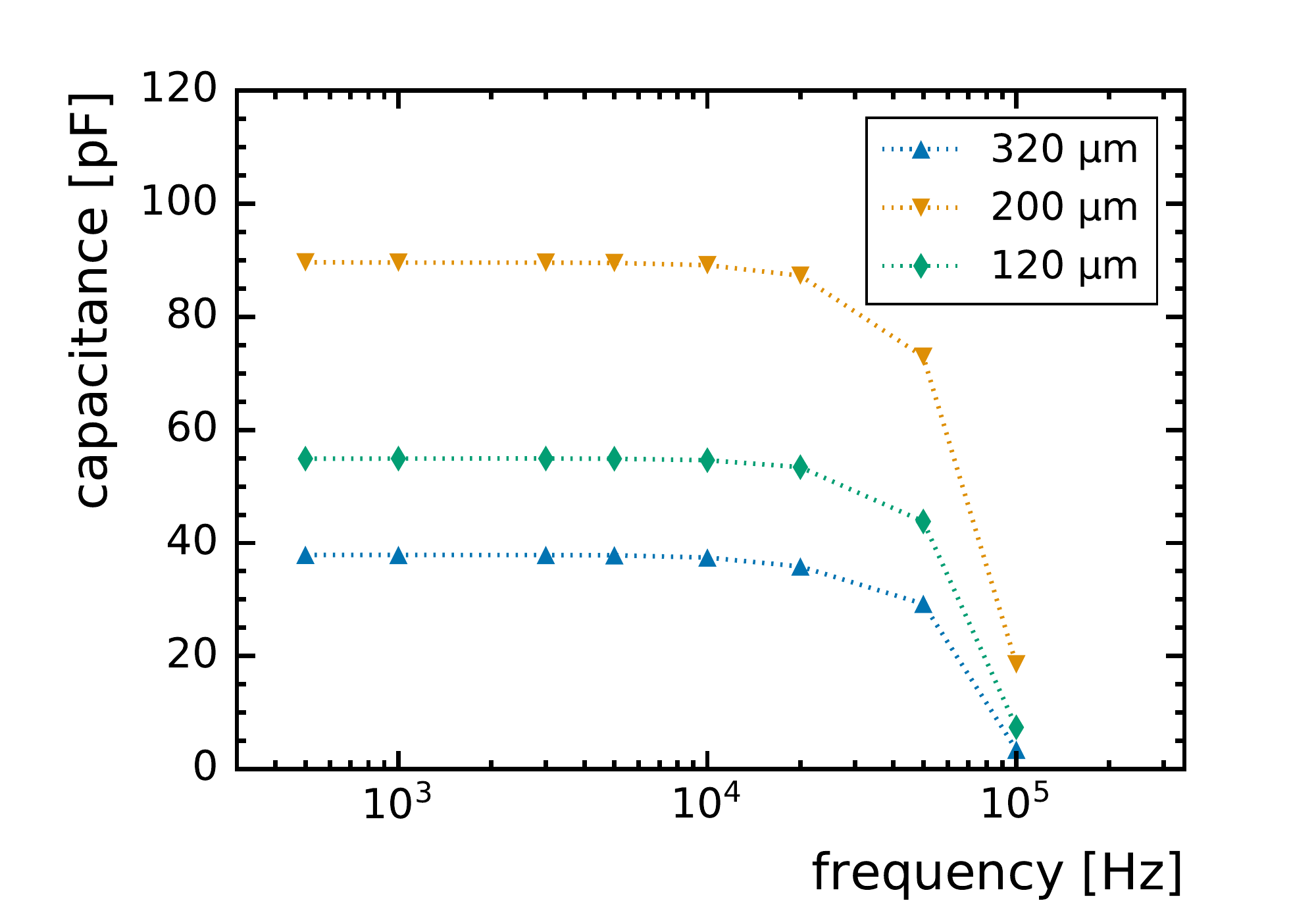}
        \label{fig:cv_freqs}}
    \caption{(a) CV curves for a set of three example sensors with identical frontside processing but different active thickness. Measured with 10\,kHz. (b) The frequency dependency of the measured capacitance at the maximum voltage point.}
\end{figure}

Fig.~\ref{fig:cv_curve} shows the corrected CV curves measured at 10\,kHz and R$_\text{bias}$ set to 10\,M\textOmega~from a set of 8-inch n-on-p HGCAL prototype sensors that have been produced in 120\,\textmu m, 200\,\textmu m and 320\,\textmu m active thickness with identical frontside processing.
The deviation of the measured capacitance values for identical pads across a full sensor is typically around 0.1\,pF.
Fig.~\ref{fig:cv_freqs} shows the measured capacitance values of the same sensors at the maximum voltage point from Fig.~\ref{fig:cv_curve} as a function of the measurement frequency. The measurement results are stable up to 10\,kHz.
At higher frequencies, the series resistance of the multiplexer network prevents a correct measurement (see Sec.~\ref{sec:cv_precision}).


\section{Discussion of System Aspects}
\label{sec:discussion}


\subsection{Measurement Range}
For the current measurement, the lower limit is given by the precision of the external measurement device that is used. Currents of a few hundred pA have been routinely measured.
The upper limit is given by the voltage drop across the series resistance of the multiplexer network R$_\text{series}$. If a drop of 1\,V is acceptable, the limit is about 100\,\textmu A.
Pads with higher currents, e.g. pads with breakdown voltages lower than the design value, are masked in software to avoid further measurements at higher voltages of this specific pad.
The pad is then shorted to ground at all times.
For the capacitance measurement, values between 5 and 100\,pF have been tested.


\subsection{Measurement of Irradiated Sensors}
For irradiation studies going beyond hadron fluences of 10$^{15}$ 1\,MeV neutron equivalent per cm$^2$, as will be the case for HL-LHC, the leakage current of the sensors will exceed 1\,\textmu A/cm$^2$, even when cooled to temperatures of -30$^\circ$C.
The total current for an 8-inch sensor will likely reach a few mA.
For such studies, the voltage drop across R$_\text{HV}$ (see Fig.~\ref{fig:schematic}), which has to carry the full sensor current, becomes non-negligible.
In the default implementation, this value is set to 40\,k\textOmega.
Depending on the desired total current range and acceptable voltage drop, this value can be reduced without significant impact on any measurement.
If simultaneous CV studies are desired for pads with large per-cell currents, R$_\text{bias}$ becomes the limiting resistor.
For safe operation of the multiplexers, the voltage drop across R$_\text{bias}$ should not exceed 10\,V.
The value of this resistor can be set in software to values between 100\,k\textOmega~and 100\,M\textOmega.
The effect on the precision of the capacitance measurement depends on the capacitance and frequency of interest, as discussed in the following section.

\subsection{Precision of the Capacitance Measurement}
\label{sec:cv_precision}
The two main factors determining the precision of the capacitance measurement are R$_\text{bias}$, which increases the impedance of the circuit parallel to the pad-under-test, and the series resistance of the multiplexer network R$_\text{series}$.
To estimate their influence, a SPICE simulation~\cite{ref:spice} of the system including 270 parallel sensor pads, corresponding to a specific probe card layout, has been implemented.
Parasitic capacitances to ground and to neighbouring channels have been added to the simulation according to laboratory measurements.

\begin{figure}[thb]
\begin{center}
    \subfigure[]{
        \includegraphics[width=0.49\textwidth]{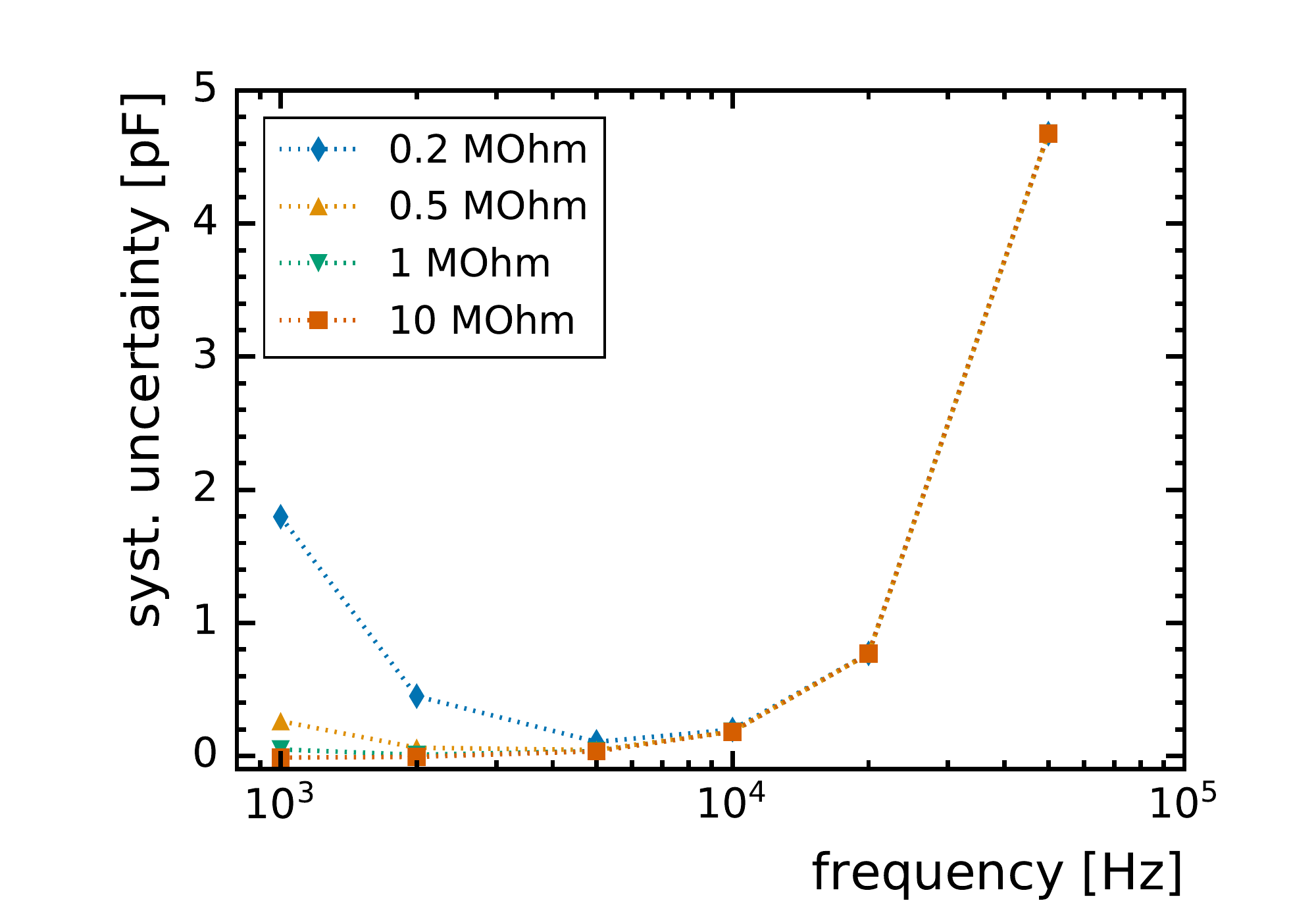}
        \label{fig:precision_vs_freq}}%
    \subfigure[]{
        \includegraphics[width=0.49\textwidth]{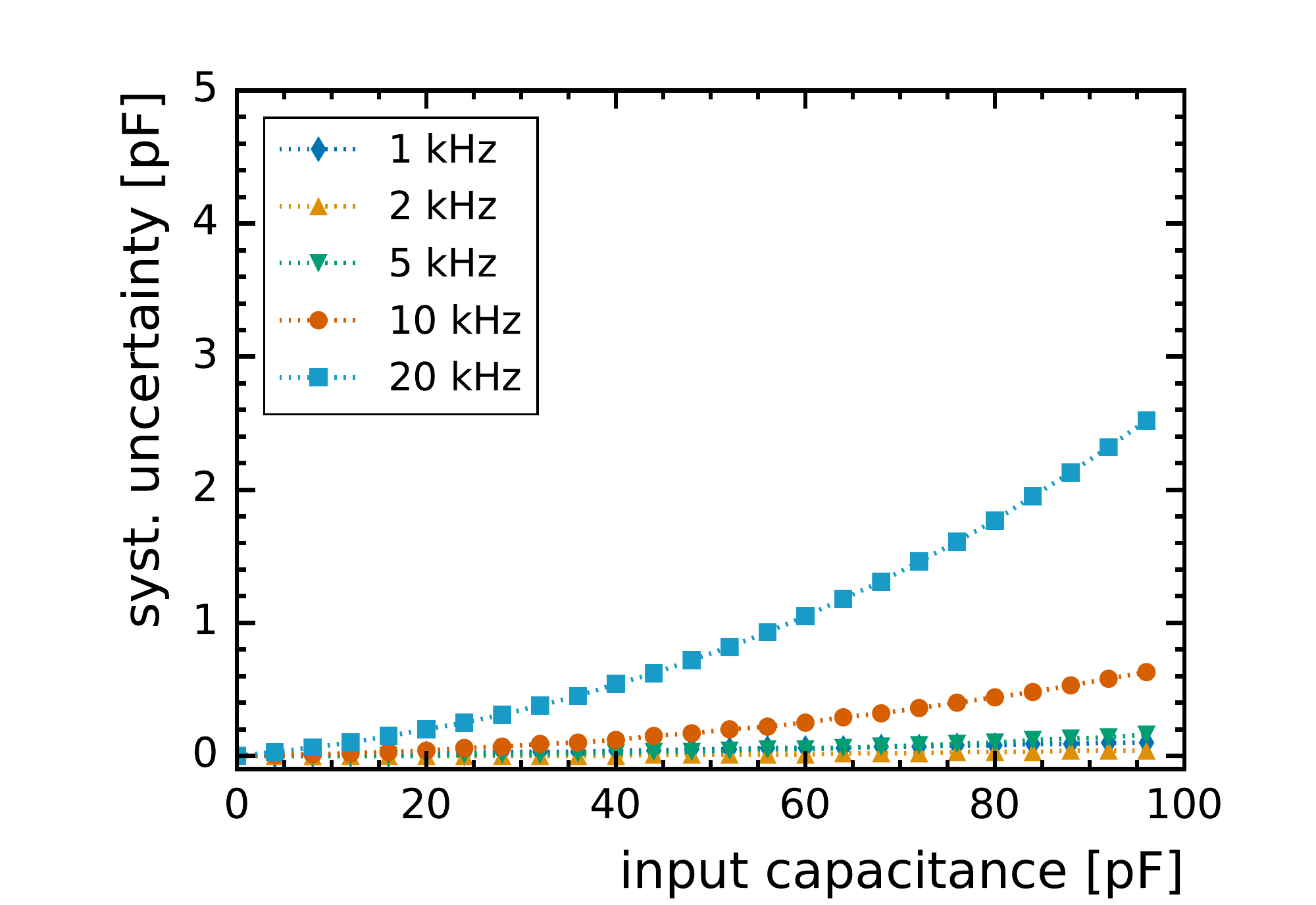}
        \label{fig:precision_vs_cap}}
    \caption{Predicted systematic uncertainty of the capacitance measurement from SPICE simulations for (a) different values of R$_\text{bias}$ at a pad capacitance of 50\,pF and (b) at R$_\text{bias}$\,=\,1\,M\textOmega~for different frequencies as a function of the pad capacitance.}
    \label{fig:cv_precision}
\end{center}
\end{figure}

Fig.~\ref{fig:precision_vs_freq} shows the systematic measurement uncertainty as a function of the measurement frequency at a pad capacitance of 50\,pF for different values of R$_\text{bias}$.
At low frequencies, the pad impedance becomes comparable to an O(1\,M\textOmega) resistor.
If the value of R$_\text{bias}$ is too small, the impedance of the circuit parallel to the pad-under-test is not negligible anymore.
At higher frequencies, the pad impedance decreases and eventually becomes comparable to O(1\,k\textOmega) and therefore R$_\text{series}$.
In this case, the parallel equivalent model used to calculate a capacitance from the measured impedance becomes invalid and introduces a systematic uncertainty.
While R$_\text{series}$ is a given limitation of the system, R$_\text{bias}$ can be set in software according to the measured leakage current.
Fig.~\ref{fig:precision_vs_cap} shows the systematic uncertainty on the measurement as a function of the pad capacitance for different frequencies at R$_\text{bias}$\,=\,1\,M\textOmega.
A systematic uncertainty of better than 0.2\,pF can be maintained over the range from 1 to 100\,pF with any frequency between 1 and 5\,kHz.

\section{Summary}
\label{sec:summary}

ARRAY is a compact, modular, precise and cost efficient system for large area silicon pad sensor characterisation.
The system consists of two plugin PCBs: an active switching matrix with 512 input channels that holds all controls and a passive probe card that connects to the sensor.
The latter can be adapted to any sensor geometry.
Six such cards have been designed so far for the CMS and FCAL collaborations.
The system delivers accurate measurements for a large range of cell currents and capacitances. The maximum bias voltage that can be applied to the sensor is $\pm$1\,kV.
Currents ranging from 500\,pA to 5\,\textmu A and capacitances between 5\,pF and 100\,pF have been measured so far.
A precision of better than 0.2\,pF on capacitance measurements in that range can be achieved.
The design files, firmware and two software implementations are available under open source licenses.

\section*{Acknowledgements}
The authors would like to thank the CMS collaboration, especially the CMS HGCAL silicon sensor group, and the FCAL collaboration for providing the sensors for these tests. In particular, we would like to thank Zoltan Gecse, Ron Lipton and Paul Rubinov (all Fermilab National Accelerator Lab., US) who first developed the concept of using spring-loaded pins to contact the HGCal sensor and designed a first version of the probe card as well as Itamar Levy and Meny Raviv Moshe (both Tel Aviv University, Israel) for their help with the LumiCal measurements. The CLICdp collaboration we thank for advice and support. We are also grateful for the help from the CERN PCB design office and the CERN SMD workshop in design and construction of the two card system as well as Fernando Duarte Ramos' (CERN) support with the mechanical integration. This project has received funding from the Austrian Doctoral Student Programme at CERN.


\section*{References}

\bibliographystyle{elsarticle-num}
\bibliography{bibliography}

\end{document}